\documentclass{kapproc} 

\usepackage{procps} 

\usepackage[dvips]{graphicx}

\upperandlowercase

\kluwerbib 

\def\apj{ApJ}           
\def\apjl{ApJ}          
\def\aap{A\&A}          
\def\mnras{MNRAS}       

\def\oiii{[O\,{\sc iii}]}
\def\halpha{\hbox{H$\alpha$}}
\def\hbeta{\hbox{H$\beta$}}

\def\kms{$\mbox{km s}^{-1}$}
\newcommand{\sauron}{{\texttt {SAURON}}}

\begin{document}

\articletitle{A \sauron\ study of stars and gas in Sa bulges}
\articlesubtitle{}

\author{J. Falcon-Barroso\altaffilmark{1*}, R. Bacon\altaffilmark{2}, M.
Bureau\altaffilmark{3}, M. Cappellari\altaffilmark{1}, R.~L.
Davies\altaffilmark{3}, P.~T. de Zeeuw\altaffilmark{1}, E.
Emsellem\altaffilmark{2}, K. Fathi\altaffilmark{4}, D.
Krajnovic\altaffilmark{3}, H. Kuntschner\altaffilmark{5}, R.~M.
McDermid\altaffilmark{1}, R.~F. Peletier\altaffilmark{6},
M. Sarzi\altaffilmark{3}}

\altaffiltext{1}{Sterrewacht Leiden, Niels Bohrweg~2, 2333~CA Leiden, The
Netherlands}
\altaffiltext{2}{CRAL - Observatoire, 9~Avenue Charles Andre, 69561 Saint
Genis Laval, France}
\altaffiltext{3}{University of Oxford, Keble Road, OX1 3RH, Oxford, United
Kingdom}
\altaffiltext{4}{Rochester Institute of Technology, 84 Lomb Memorial Dr,
Rochester, NY 14623-5603, USA}
\altaffiltext{5}{European Southern Observatory, Karl-Schwarzschild-Str~2, 85748
Garching, Germany}
\altaffiltext{6}{Kapteyn Astronomical Institute, P.O. Box 800, 9700 AV
Groningen, The Netherlands}

\email{$^{*}$jfalcon@strw.leidenuniv.nl}

\begin{abstract}
We present results from our ongoing effort to understand the morphological and
kinematical properties of early-type galaxies using the integral-field
spectrograph \sauron. We discuss the relation between the stellar and gas
morphology and kinematics in our sub-sample of 24 representative Sa spiral
bulges. We focus on the frequency of kinematically decoupled components and on
the presence of star formation in circumnuclear rings.
\end{abstract}

\section{Sa bulges in the \sauron\ Survey}
The starting point of our study is the \sauron\ survey of nearby, early-type
galaxies (de Zeeuw et al. 2002). The full sample consists of 72 galaxies
divided in three morphological groups: 24 ellipticals, 24 lenticulars, and 24 Sa
bulges. Within each group, half of the galaxies are drawn from the field
environment, and the remaining half are representative of the cluster
population.

The main science driver for the survey is the study of the intrinsic shapes,
velocity and metallicity distribution, and the relation between the stellar and
gas kinematics to the underlying stellar populations. Most of the results of the
survey up to now have focused on the 48 elliptical and lenticular galaxies (de
Zeeuw et al. 2002, Emsellem et al. 2004, Cappellari et al. 2005, Sarzi et
al. 2005). Here, we concentrate on the 24 Sa spiral bulges.

A key question we want to address is whether secular evolution plays an
important role in the formation of bulges (i.e. disk instabilities, Pfenniger \&
Norman 1990, Pfenniger \& Friedli 1991), or whether they are scaled-down
versions of ellipticals, as the main scaling relations suggest (e.g.,
Fundamental Plane, Jorgensen et al. 1996, Falcon-Barroso et al. 2002). A
comprehensive description of the stellar and gas kinematics of the complete
sample of 24 Sa spiral bulges is presented in Falcon-Barroso et al. (2005). We
highlight some of the results here. Fathi et al. (2005) present an in depth
study of an individual case (NGC\,5448). Ganda et al. (2005) discuss similar
\sauron\ observations of 18 later-type spiral galaxies.

\section{Kinematically Decoupled Components}
The stellar kinematic maps of our galaxies display a variety of substructures in
the central regions, such as kinematically decoupled components. For example,
Figure~1 (top panel) shows the stellar kinematics of NGC\,5689. In that galaxy
the component is clearly visible in the velocity map as a pinching of the
isovelocities in the inner regions, which is also accompanied by a decrease of
the velocity dispersion and an anti-correlated $h_3$ parameter at the same
locations. In general this type of decoupled components appear to be aligned
with the major axis of the galaxy and tend to be associated with dust disks or
rings seen in unsharped masked images. However, there are a few cases in our
sample where the rotation axis of the decoupled component is misaligned with
respect to the main rotation axis of the galaxy (e.g., NGC\,4698, see Figure~1
bottom panel). In this case the decoupled component can be identified as a twist
of the zero velocity curve in the inner regions. We find that 13 out of 24
galaxies (52\%) display clear signatures of such structures. Despite the high
number of kinematically decoupled components observed, the true fraction can be
even higher as their detection can be hampered by inclination (i.e., as it is
more difficult to detect them at low galaxy inclinations), but also by a limited
spatial sampling.

\begin{figure}[!ht]
\includegraphics[width=0.99\linewidth]{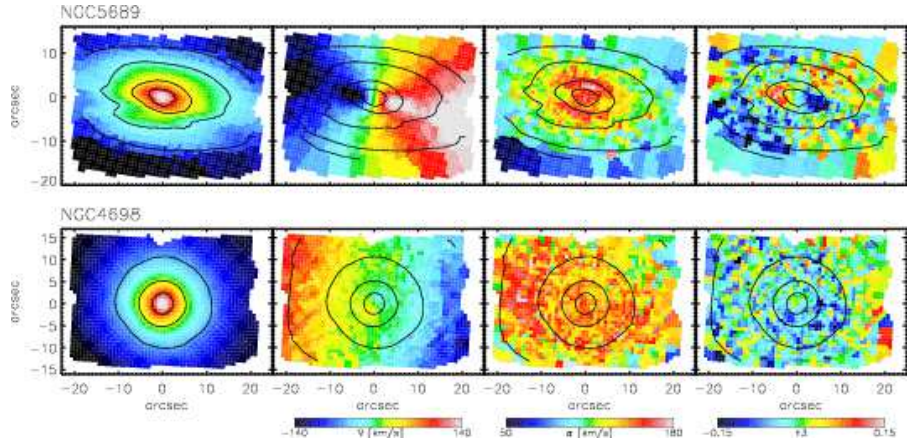}
\caption{Stellar kinematic maps for NGC\,5689 and NGC\,4698. In each column
(from left to right) we show the reconstructed intensity image from the \sauron\
datacube (in mag/arcsec$^2$), the stellar radial velocity, velocity
dispersion (in \kms) and $h_3$ Gauss-Hermite moment.}
\end{figure}

The nature of these kinematically decoupled components is still uncertain, and
there are different formation and evolution processes that can produce such a
component in the inner regions of the galaxies. One of the most widely supported
scenarios is that these objects formed by gas inflow towards the central regions
of the galaxy which consequently induced star formation (Wozniak et al. 2003).
The role of bars and dissipative processes, however, is not yet well understood
in this context (Bureau \& Athanassoula 2005, Heller \& Shlosman 1994,
Friedli \& Benz 1995), as all these models are able to reproduce the observed
velocities, inwards decrease of velocity dispersion and the $h_3$
anti-correlation. Under this scenario the kinematically decoupled component is
formed from existing material in the galaxy.

An alternative possibility for the formation of these components is an
interaction event. This model is preferred to explain the presence of misaligned
components such as the one seeing in NGC\,4698. This galaxy is a well-studied
case in which the formation of the component is thought to be the result of an
intermediate merger event (Bertola et al. 1999). In Figure~2, we present the
case of NGC\,5953, which offers a unique opportunity to study the effects that
an on-going interaction has on the stellar and gas properties. The kinematical
decoupling in the inner parts, together with the presence of large amounts of
fresh gas, suggests that we might be witnessing the early stages in the
formation of a kinematically decoupled component. 

\begin{figure}[!ht]
\includegraphics[width=0.99\linewidth]{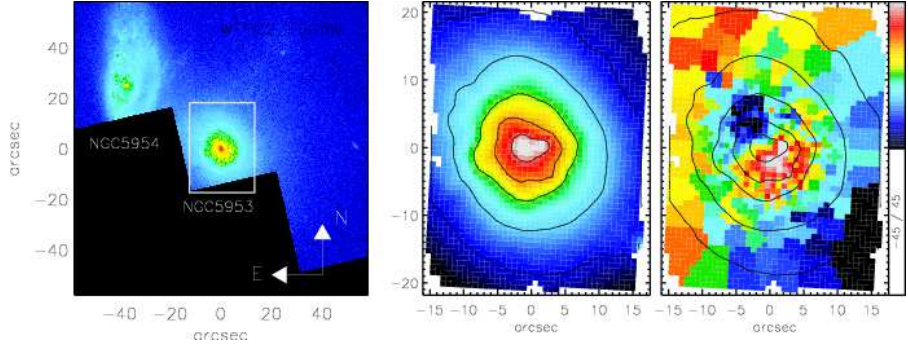}
\caption{Interacting pair NGC\,5953/NGC\,5954. From left to right: i) HST/WFPC2
image of the pair of galaxies. Overlaid on NGC\,5953 is the \sauron\
field-of-view. ii) \sauron\ reconstructed intensity image of NGC\,5953, iii)
stellar radial velocity of NGC\,5953.}
\end{figure}

\section{Global and circumnuclear star formation}
Several methods have been used in the past to trace star formation (SF) in
galaxies. Given the little observational overhead, the emission of Balmer
lines has become one of the most popular tracers (Kennicutt 1998). In addition
to the Balmer lines, the \oiii/\hbeta\ ratio can also serve as a diagnostic in
situations where SF is intense and derives from pre-enriched material (Kauffmann
et al. 2003). Here we use both diagnostics to investigate the importance of
star formation in Sa galaxies.

The morphology of the star-forming regions in our sample appears to have
multiple forms, although as expected in spiral galaxies, the bulk of SF is
concentrated in the main disk. Within galaxies, we find that SF activity is more
intense along dust lanes, has an amorphous morphology, or is confined in
circumnuclear rings.

These rings represent one of the most spectacular forms of SF. The frequency of
these circumnuclear regions in our sample is in good agreement with the most
recent studies from \halpha\ observations of larger samples (21$\pm$5\%, Knapen
2005). In Figure~3, we present the cases of NGC\,4245, NGC\,4274, and NGC\,4314
to illustrate the most common properties of the stars and gas in these SF
regions. There is a remarkable correlation between the dust and the regions
where SF is very intense (i.e., large \hbeta\ flux). At the same time, the
\oiii/\hbeta\ ratio is very low at the same locations, confirming the presence
of star formation. The velocity dispersion of the ionised gas is also low
($\sim$50 \kms) along the star-forming rings. We are thus seeing stars being
formed from dynamically cold gas. A similar behaviour is also found in other
galaxies in the sample with extreme SF activity.

\begin{figure}[ht]
\includegraphics[angle=0,width=0.99\linewidth]{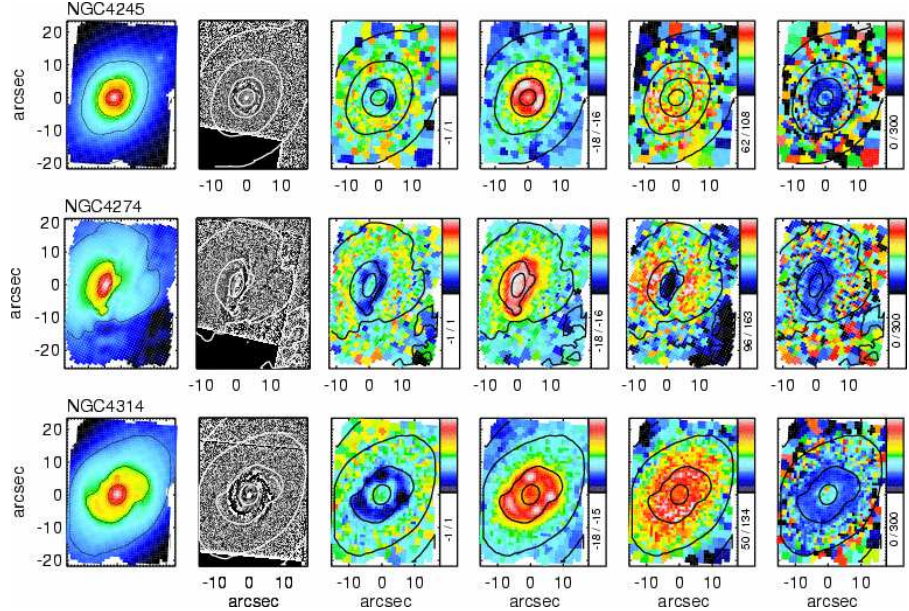}
\caption{Stellar and gas maps for three star-forming ring galaxies NGC\,4245,
NGC\,4274, NGC\,4314. In each column (from left to right) we show i) the
reconstructed intensity image from the \sauron\ datacube, ii) an HST
unsharp-masked image, the iii) \oiii/\hbeta\ ratio (logarithmic scale), \hbeta\
flux (in units of erg/cm$^2$ s$^{-1}$, and logarithmic scale), iv) stellar velocity
dispersion (in \kms), and v) ionised-gas velocity dispersion (in \kms).}
\end{figure}

As discussed in Section~2, star formation can produce the observed stellar
velocity dispersion drops (also called `sigma-drops'). In our sample we find
that the presence of young stars does not necessarily relate directly to the
stellar velocity dispersion ($\sigma_{*}$). In the examples shown in Figure~3,
only NGC\,4274 displays a sigma-drop at the location of the young stars.
NGC\,4245 shows a small decrease  in $\sigma_{*}$ only on the brightest \hbeta\
knots along the ring. In NGC\,4314, one of the galaxies with highest apparent SF
activity, the presence of young stars hardly affects $\sigma_{*}$. In practice,
these features strongly depend not only on the number of young stars present,
but also on the inclination of the galaxy, as in face-on configurations (e.g.,
NGC\,4314) the contribution of the young stars to the line-of-sight radial
velocity is much smaller than that of the surrounding bulge.

The formation of star-forming rings is generally associated with bar-driven gas
inflow towards the inner regions of galaxies (i.e., gas material gets trapped 
in rings as it approaches an important resonance). SF in the ring then begins 
once the gas density in the ring exceeds the limit set by the Toomre (1969) 
criterion. In our sample only three of the six galaxies with star-forming rings
are known to be barred (NGC\,4245, NGC\,4274, NGC\,4314). For the remaining
three galaxies with rings (NGC\,2844, NGC\,5953, NGC\,7742), the interacting
scenario offers an alternative explanation (see Figure~2). Although the case of
NGC\,7742 is still unclear, as the presence of an oval (i.e., weak bar) could
also explain the observed star-forming ring (see Figure 13 in de Zeeuw et al.
2002).

\section{Conclusions}
We are carrying out a comprehensive study of the kinematical properties of a
sample of 24 Sa bulges drawn from the \sauron\ survey of representative
galaxies. We find that kinematically decoupled components are present in half of
our sample of galaxies (13/24). They are easily detected in the stellar velocity
maps, but often also leave an imprint in the stellar velocity dispersion and
$h_3$ parameters. Despite the growing evidence that they are the result of star
formation induced by bar-driven gas inflow towards the inner parts of galaxies,
interactions as well as non-dissipative bar evolution models can also reproduce
the observed kinematics.

Star formation in our sample displays different morphologies. The most striking
cases appear in the form of circumnuclear rings. We find a good correlation
between star formation and the velocity dispersion of the ionised gas in these
rings, indicating that we are seeing young stars being formed from dynamically
cold gas. The young stars in these rings often produce a decrease of the stellar
velocity dispersion, although the presence of these sigma-drops strongly
depends on the number of young stars, but more importantly on the inclination
of the galaxy. While the presence and properties of circumnuclear rings can be
easily explained in the context of bars, interactions can also reproduce the
observed morphological and kinematic properties.

\section*{Acknowledgements}
JFB acknowledges support from the Euro3D Research Training Network,
funded by the EC under contract HPRN-CT-2002-00305.

\bibliographystyle{kapalike}
\chapbibliography{logic}

\end{document}